\documentclass[lettersize,journal]{IEEEtran}
\usepackage{amsmath,amsfonts}
\usepackage{algorithmic}
\usepackage{algorithm}
\usepackage{array}
\usepackage[caption=false,font=normalsize,labelfont=sf,textfont=sf]{subfig}
\usepackage{textcomp}
\usepackage{stfloats}
\usepackage{url}
\usepackage{verbatim}
\usepackage{graphicx}
\usepackage{cite}
\usepackage[nolist]{acronym}
\usepackage[draft]{pdfcomment}
\usepackage[dvipsnames]{xcolor}

\begin{acronym}
\acro{BioZ}{bioimpedance}
\acro{IA}{instrumentation amplifier}
\acro{SDCM}{source degenerated current mirror}
\acro{DTMOS}{dynamic threshold MOS}
\acro{TC}{transconductance}
\acro{TI}{transimpedance}
\acro{FVF}{flipped voltage follower}
\acro{LC}{lead compensation}
\acro{CMFB}{common-mode feedback}
\acro{THD}{total harmonic distortion}
\acro{IC}{integrated circuit}
\acro{VGS}{gate-source voltage}
\acro{PVT}{process voltage temperature}
\acro{LLCM}{long length current mirror}
\acro{SDCM}{source degenerated current mirror}
\acro{PSD}{power spectral density}
\acro{NEF}{noise efficiency factor}
\end{acronym}
\hypersetup{hidelinks}      

\IEEEoverridecommandlockouts
\IEEEpubid{\begin{minipage}[t]{\textwidth}\ \\[10pt]
\centering{This work has been submitted to the IEEE for possible publication.\\
Copyright may be transferred without notice, after which this version may no longer be accessible.}
\end{minipage}}

\begin{document}
\title{A 2.5$\,\mu$W 30\,nV/$\surd$Hz Instrumentation Amplifier for Bioimpedance Sensors with Source Degenerated Current Mirror and DTMOS Transistor}
\author{
Yu~Xue,~\IEEEmembership{Student Member,~IEEE,}
Kwantae~Kim,~\IEEEmembership{Senior Member,~IEEE,}
\thanks{This work was supported in part by the Aalto SemiSummer2025 Training Program granted by Aalto University School of Electrical Engineering and funded by the Finnish Semiconductor Industry.}
\thanks{The authors are with the Department of Electronics and Nanoengineering, School of Electrical Engineering, Aalto University, 02150 Espoo, Finland.}
}

\maketitle

\begin{abstract}
This paper proposes a low-power and low-noise instrumentation amplifier (IA) tailored for bioimpedance sensing applications. The design originates from a gain-boosted flipped voltage follower (FVF) transconductance (TC) stage and integrates two complementary circuit techniques to improve the noise performance. To achieve an optimal balance between input-referred noise and available voltage headroom, a source-degenerated current mirror (SDCM) is adopted, resulting in reducing the input-referred noise by 7.95\% compared with a conventional current mirror structure. In addition, a dynamic threshold MOSFET (DTMOS) scheme is employed to enhance the effective transconductance, leading to a further 11.66\% reduction in input-referred noise. Simulated in a 28\,nm CMOS process demonstrate that the proposed IA achieves an input-referred noise floor of \text{30}\,nV/$\surd\text{Hz}$ and a bandwidth of \text{1.44}\,MHz, while consuming only \text{2.5}\,$\mu$W from a 0.8\,V supply. Compared to the baseline design, the proposed approach achieves a \text{32.4\%} reduction in power consumption without degrading noise performance. The complete design parameters are open-sourced in this paper, to ensure reproducibility and facilitate future developments.
\end{abstract}


\section{introduction}
\IEEEPARstart{B}{ioimpedance} (\acs{BioZ}) sensing has emerged as a key modality for physiological monitoring, leveraging the frequency-dependent properties of biological tissues for real-time vital sign estimation. Applications span clinical and consumer domains, including heart rate~\cite{Cheon2024bioz-isscc}, respiration~\cite{Kim2017BioZ-ESSCIRC}, glucose monitoring~\cite{Ollmar2023glucose-sr, Song2015JSSC-Glucose}, hand gesture recognition~\cite{Jiang2020eit-tcasii}, and early cancer detection~\cite{Lee2020eit-cicc}. Compact wearable \cite{Cheon2024bioz-isscc, Pan2024bioz-jssc} and implantable \cite{Ollmar2023glucose-sr} devices impose demand high-fidelity \ac{BioZ} under tight power budgets, making low-power analog front-end essential due to miniaturization of battery-operated devices. This has driven extensive research into analog front-ends \cite{Cheon2024bioz-isscc, Kim2017EIT, Kim2020BioZ, Kim2017BioZ-ESSCIRC, Pan2024bioz-jssc, Kim2023BioZ-ISCAS, Teng2014BioZ-Biocas, Takhti2019bioz-tbiocas}, where \acp{IA} are key to low-noise signal readout. Advances in low-noise design are critical to meet the resolution and energy demands of wearable and implantable \ac{BioZ} systems. 

\begin{figure}[t]
    \begin{center}
    \includegraphics[width=0.95\columnwidth]{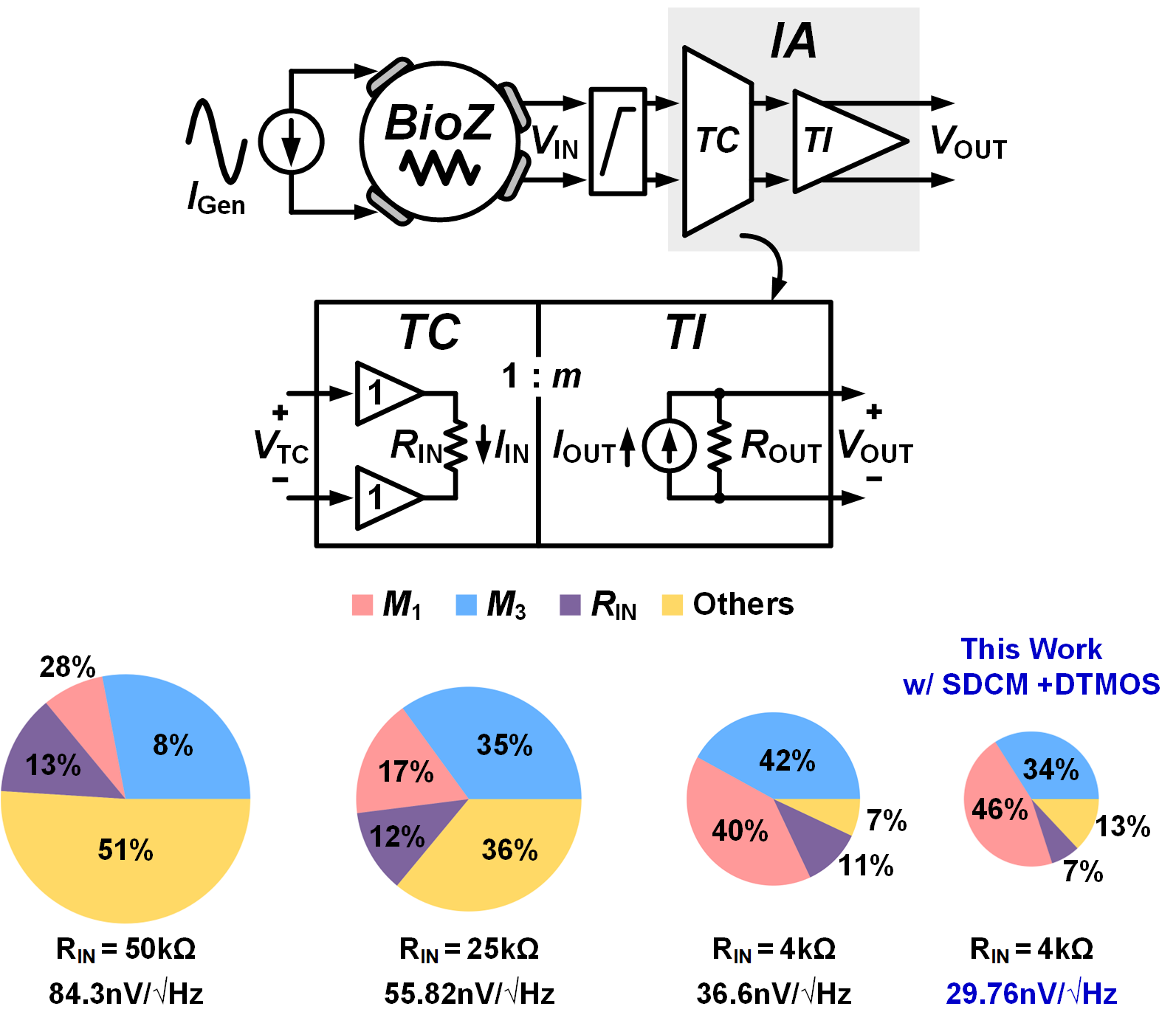}
    \caption{Typical architecture of an \ac{IA} in \ac{BioZ} voltage readout \cite{Kim2020BioZ, Kim2017EIT, Kim2023BioZ-ISCAS, Teng2014BioZ-Biocas} and the proposed \ac{IA} and their noise breakdown at 100\,kHz when simulated in 28\,nm CMOS process. $M_{1}$ and $M_{3}$ can be found in Fig.~\ref{fig:baseline}.}\label{fig:architecture_BioZ}
    \end{center}
    \vspace{-5mm}
\end{figure}

Fig.~\ref{fig:architecture_BioZ} illustrates a typical voltage-readout front-end architecture in a \ac{BioZ} sensor \ac{IC}, incorporating an \ac{IA} for signal amplification with minimized input-referred noise to detect subtle physiological variations \cite{Kim2020BioZ, Kim2017EIT, Cheon2024bioz-isscc, Lee2020eit-cicc, Kim2023BioZ-ISCAS, Pan2024bioz-jssc, Teng2014BioZ-Biocas, Takhti2019bioz-tbiocas}. The \ac{IA} comprises \ac{TC} and \ac{TI} stages, where its gain is $mR_\text{OUT}/R_\text{IN}$ \cite{Kim2023BioZ-ISCAS}, enabling precise voltage gain by resistance ratio. The \ac{TC} stage is the front-end of the readout chain, employing a \ac{FVF}-based buffer which is critical to overall noise, bandwidth, power consumption, and linearity performance, thereby directly impacting the accuracy and robustness of the \ac{BioZ} measurement. The subsequent \ac{TI} stage converts the current signal of the \ac{TC} stage into a differential voltage across $R_\text{OUT}$.

\begin{figure*}[t]
    \begin{center}
    \includegraphics[width=\textwidth]{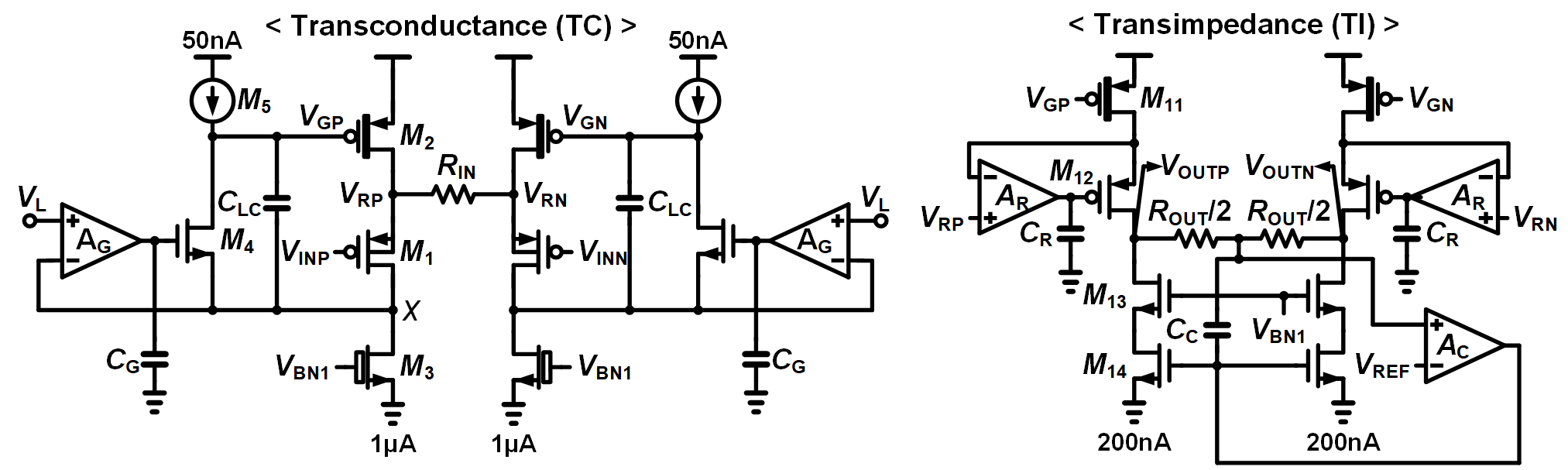}
    \caption{Schematic of the baseline \ac{IA} \cite{Pan2022BioZ,Zhang2023BioZ-JSSC,Kim2020BioZ,Kim2023BioZ-ISCAS}.}\label{fig:baseline}
    \end{center}
    \vspace{-3mm}
\end{figure*}

\begin{figure*}[t]
    \begin{center}
    \includegraphics[width=2\columnwidth]{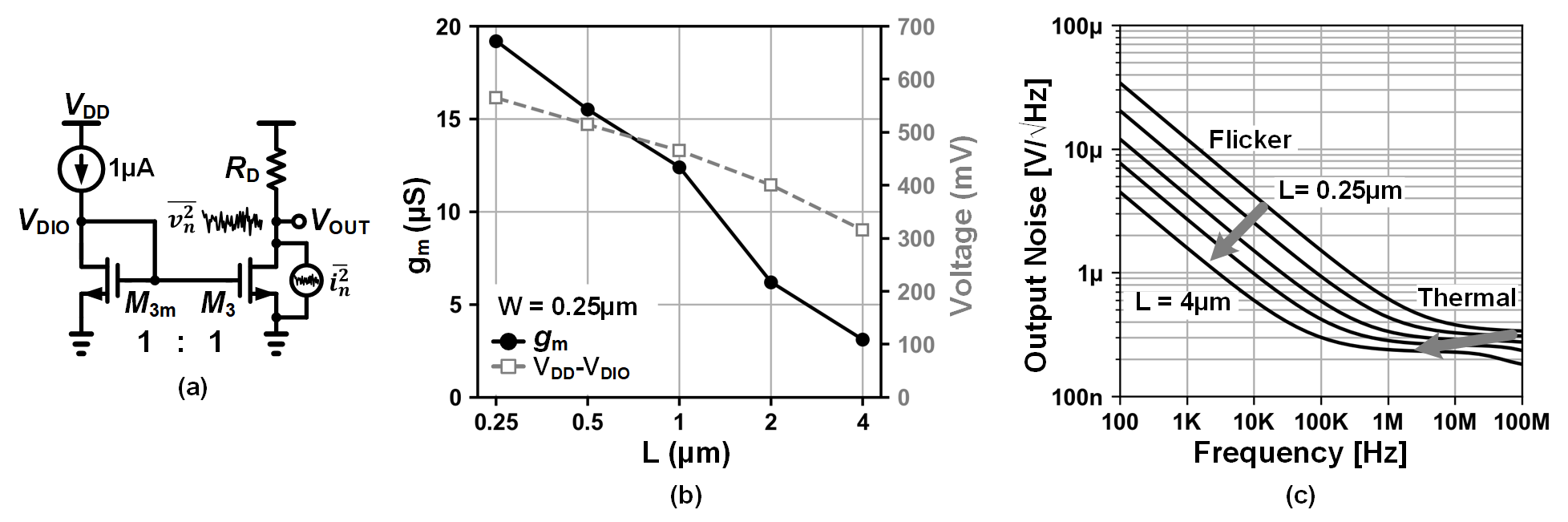}
    \caption{(a) Schematic of a conventional current mirror, (b) simulated $g_m$ and voltage headroom with different lengths, and (c) output noise spectral density versus frequency for different lengths.}\label{fig:LLCM}
    \end{center}
    \vspace{-5mm}
\end{figure*}

Prior works \cite{Kim2017EIT, Kim2023BioZ-ISCAS, Kim2020BioZ, Takhti2019bioz-tbiocas} have identified the input resistor $R_\text{IN}$ as one of the primary contributors to the input-referred noise of \ac{IA}. To achieve low noise performance, a small $R_\text{IN}$ is preferred, while the associated loop gain degradation in the \ac{FVF} circuit can be compensated by gain enhancement techniques. The input-referred thermal noise \ac{PSD} of the \ac{IA} can be expressed as\cite{Zhang2023BioZ-JSSC}:
\begin{align} \label{eq:noise_IA_baseline}
    \overline{v_{n,\text{IA}}^{2}}
    &\approx \overline{v_{n,\text{M1}}^{2}}
    + \left( \overline{i_{n,\text{M3}}^{2}} + \overline{i_{n,R_\text{IN}/2}^{2}} \right)
    \left(\frac{R_\text{IN}}{2}\right)^{2} \notag\\
    &= \frac{4kT\gamma}{g_{m1}}
    + \left(
        4kT\gamma g_{m3} + 4kT\frac{R_\text{IN}}{2}
    \right)
    \left(\frac{R_\text{IN}}{2}\right)^{2}
\end{align}
where $M_1$ and $M_3$ denote the input transistor and the tail current source in the \ac{TC} stage, respectively (Fig.~\ref{fig:baseline}), $g_{m}$ is the transistor transconductance, $k$ is the Boltzmann constant, and $\gamma$ is the noise model parameter. As illustrated in Fig.~\ref{fig:architecture_BioZ}, reducing $R_\text{IN}$ from 50\,\text{k}$\Omega$ to 4\,\text{k}$\Omega$ significantly suppresses the input referred noise and make the noise contribution of $R_\text{IN}$ to a negligible level. However, the noise contribution from $M_3$ remain significant, accounting for approximately 42\% of the total input-referred noise and forming a bottleneck to further improvement in noise performance.

Beyond noise performance, voltage headroom has also become a critical design aspect of analog circuits in advanced CMOS processes. As the supply voltage $V_\text{DD}$ continues to scale down with technology, the available margin for signal swing and biasing is reduced \cite{Nauta2024ISSCC, Kinget2015ISSCC, Kim2022KWS-JSSC}. This reduced headroom poses challenges for the design of \acp{IA} and is essential to address in parallel with noise reduction to ensure robust performance.

\begin{figure*}[t]
  \centering
  \includegraphics[width=\textwidth]{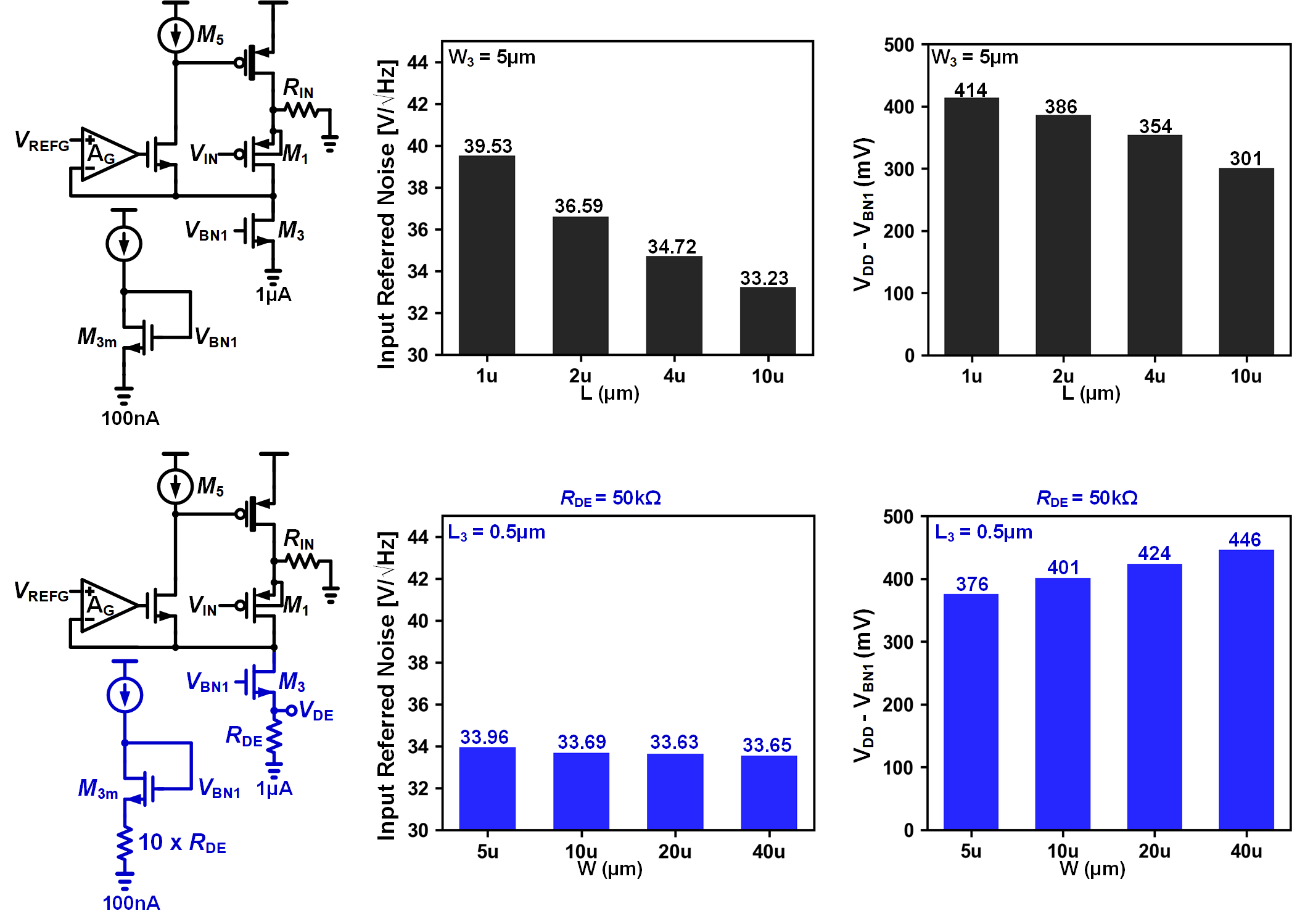}
  \caption{Differential-mode half-circuit implementation and simulated performance comparison of current mirrors in the IA. The black bar plots represent the input-referred noise and voltage headroom versus transistor length $L_{3}$ using a conventional current mirror, while the blue bar plots represent the input-referred noise and voltage headroom versus width $W_{3}$ with a fixed $R_\text{DE}=50\,\text{k}\Omega$ using the proposed SDCM.}
  \label{fig:half_circuit}
\end{figure*}

In this paper, we propose applying two noise reduction techniques to the \ac{IA} for ultra-low-power \ac{BioZ} sensor ICs: 1) a \ac{SDCM} and 2) a \ac{DTMOS}. The \ac{SDCM} reduces noise from the tail current source $M_{3}$, with only a minimal penalty of 50\,mV in the voltage headroom of the amplifier core ($V_\text{DE}$ in Fig.~\ref{fig:half_circuit}). The \ac{DTMOS} is employed to lower the noise of the input transistor $M_{1}$, thereby further improving the noise performance of the \ac{IA}. Notably, the two techniques are orthogonal and complementary, enabling their combined use to achieve improvements unattainable by either technique alone. Simulation results in a 28\,nm CMOS process show that, when both schemes are applied together, the proposed \ac{IA} achieves an 18.7\% reduction in input-referred noise without any increase in power consumption compared to the baseline design. Under iso-noise conditions, the \ac{IA} shows a 32.4\% higher power efficiency.

This article is an extension of \cite{Yu2025Newcas}. The remaining part of this article is organized as follows. Section\,\ref{sec:current_mirror} provides a comparative analysis and simulation results of two current mirrors, a conventional current mirror and a \ac{SDCM}. Section\,\ref{sec:dtmos} introduces the \ac{DTMOS} scheme and its efficacy in enhancing the noise of the \ac{IA}. Section\,\ref{sec:proposed IA} presents the simulation results based on the proposed \ac{IA}. Section\,\ref{sec:conclusion} concludes this work and summarizes key contributions.

\IEEEpubidadjcol

\begin{figure*}[t!]
    \begin{center}
    \includegraphics[width=2\columnwidth]{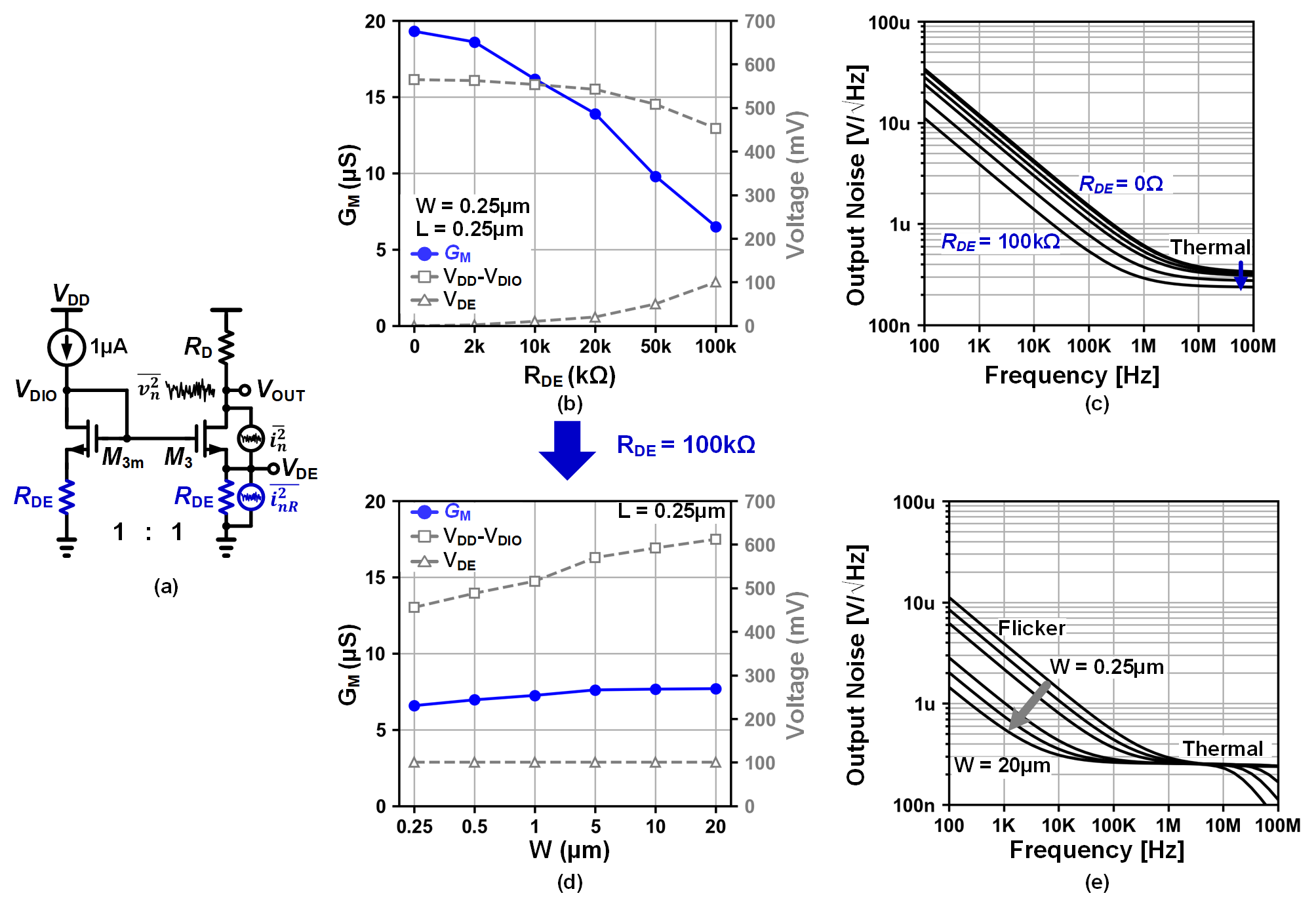}
    \caption{(a) Schematic of a \ac{SDCM}, (b) simulated $G_\text{m}$ and voltage headroom versus $R_\text{DE}$, (c) output noise spectral density versus frequency for different $R_\text{DE}$, (d) $G_\text{m}$ and voltage headroom versus width with a fixed $R_\text{DE}=100\,\text{k}\Omega$, and (e) output noise spectral density versus frequency for different width.}\label{fig:SDCM}
    \end{center}
    \vspace{-5mm}
\end{figure*}

\section{noise analysis in current mirror}\label{sec:current_mirror}

As indicated by Eq.~\eqref{eq:noise_IA_baseline}, reducing the total input-referred noise requires maximizing $g_{m1}$ while minimizing $g_{m3}$. The term $g_{m3}$, corresponding to the current mirror transistor $M_3$ in Fig.~\ref{fig:baseline}, contributes to the second term in Eq.~\eqref{eq:noise_IA_baseline}.

\subsection{Long Length Current Mirror}
Fig.~\ref{fig:LLCM}(a) illustrates a conventional current mirror, with transconductance $g_{m}$ and the thermal noise \ac{PSD} at the drain node given by
\begin{align} \label{gm}
  g_{m}
  = \frac{\partial I_\text{D}}{\partial V_\text{GS}}
  = \sqrt{ \mu C_{\text{ox}} \frac{W}{L} I_\text{D} }
  = \frac{2 I_\text{D}}{V_\text{GS} - V_\text{TH}}
\end{align}

\begin{equation} \label{eq:output_noise_Fi2.(a)}
    \overline{v_{n,\text{CM}}^2} = 4kT\gamma g_{m3} R_\text{D}^2 + \overline{v_\text{RD}^2}.
\end{equation}
where $\overline{v_\text{RD}^2}$ represents the noise \ac{PSD} contributed by $R_\text{D}$. Here, we only consider the noise contributions from the mirrored-branch, $M_{3}$ and $R_\text{D}$, and exclude the noise from $M_\text{3m}$, as the \ac{IA} is implemented in a differential architecture (Fig.~\ref{fig:baseline}). The first term in Eq.~\ref{eq:output_noise_Fi2.(a)} reveals that the noise scales with $g_{m3}$. Under strong inversion and saturation, an increase in channel length leads to a lower $g_{m}$, which consequently reduces noise~\cite{Kim2023BioZ-ISCAS, razavi2001design}. However, a lower $g_{m}$ demands a higher $V_\text{GS}$, increasing the drain voltage $V_\text{DIO}$ and degrading the voltage headroom. This constraint becomes critical in advanced CMOS processes, where the supply voltage is limited to sub-1\,V levels, as in the 28\,nm CMOS process.

Fig.~\ref{fig:LLCM}(b), (c) illustrate the trade-off between noise performance and voltage headroom by sweeping the length $L$ of the current mirror device, while keeping the width $W$ fixed at 250\,nm, based on simulations in a 28\,nm CMOS process. Since the maximum available channel length is limited to 1\,$\mu$m in the used process, we adopted a series-stacking technique~\cite{JSSC06Arnaud}, where multiple identical transistors are connected in series so that the effective channel length equals the sum of the individual lengths. As $L$ increases from 0.25\,$\mu$m to 4\,$\mu$m, $g_{m}$ decreases from 18\,$\mu$S to 3\,$\mu$S, resulting in reduced output noise. However, the accompanying increase of $V_\text{GS}$ significantly degrades the voltage headroom ($V_\text{DD}$-$V_\text{DIO}$), resulting in a 220\,mV drop from 550\,mV to 310\,mV (which is below half $V_\text{DD}$). As a result, the limited headroom makes the conventional current mirror less favorable in low-supply \ac{IA} design. This headroom problem becomes exacerbated as the CMOS process scales, since the supply voltage continues to shrink in advanced nanometer technology nodes, motivating the exploration of alternative current mirror topologies.

\subsection{Source Degenerated Current Mirror}

As shown in Fig.~\ref{fig:SDCM}(a), the transconductance of a transistor can be effectively decreased by placing a degeneration resistor in series with the source terminal. The effective transconductance $G_{m}$ and the corresponding output noise are expressed below, where the approximation holds under the condition $g_{m3}R_\text{DE} \gg 1$.
\begin{align} \label{eq:Effective GM}
    G_{m,\text{M3}} =\frac{\partial I_\text{D}}{\partial V_\text{G3}}
    = \frac{1}{\mathstrut 1/g_{m3} + R_\text{DE}}
    \approx \frac{1}{R_\text{DE}}
\end{align}

\begin{align} \label{eq:output_noise_Fi2.(b)}
    \overline{v_{n,\text{SDCM}}^2}
    &= \bigg( \frac{\overline{i_{\text{M3}}^2} + (g_{m3} R_\text{DE})^2 \overline{i_{\text{DE}}^2}}{(1+g_{m3} R_\text{DE})^2} \bigg)R_\text{D}^2 + \overline{v_\text{RD}^2} \nonumber \\
    &= \bigg( \frac{4kT\gamma g_{m3} + (g_{m3} R_\text{DE})^2 4kT/R_\text{DE}}{(1+g_{m3} R_\text{DE})^2} \bigg)R_\text{D}^2 \nonumber\\
    &\quad + \overline{v_\text{RD}^2} \nonumber\\
    &\approx \frac{4kT}{R_\text{DE}} \bigg( \frac{\gamma}{g_{m3}R_\text{DE}} + 1 \bigg)R_\text{D}^2
    + \overline{v_\text{RD}^2}
\end{align}

The complete derivation of Eq.~(\ref{eq:output_noise_Fi2.(b)}) is provided in the Appendix. Eq.~(\ref{eq:Effective GM}) shows that adding a degeneration resistor can directly reduce the effective transconductance of $M_{3}$, which is $G_{m,\text{M3}}$. Eq.~(\ref{eq:output_noise_Fi2.(b)}) indicates that the noise contribution of $M_{3}$, which is $\overline{i_\text{M3}^2}=4kT\gamma g_{m3}$, can be reduced by a factor of $(1+g_{m3} R_\text{DE})^2$ with an additional noise source from $R_\text{DE}$, $\overline{i_\text{DE}^2}=4kT / R_\text{DE}$. This reduction by the source degeneration comes from a local negative feedback similar to the source follower \cite{Kim2023filt-casm}, reducing the effective transconductance $G_{m3}=g_{m3}/(1 + g_{m3}R_\text{DE})$, thereby leading to a decreased output noise. 

\begin{figure}[t]
    \begin{center}
    \includegraphics[width=\columnwidth]{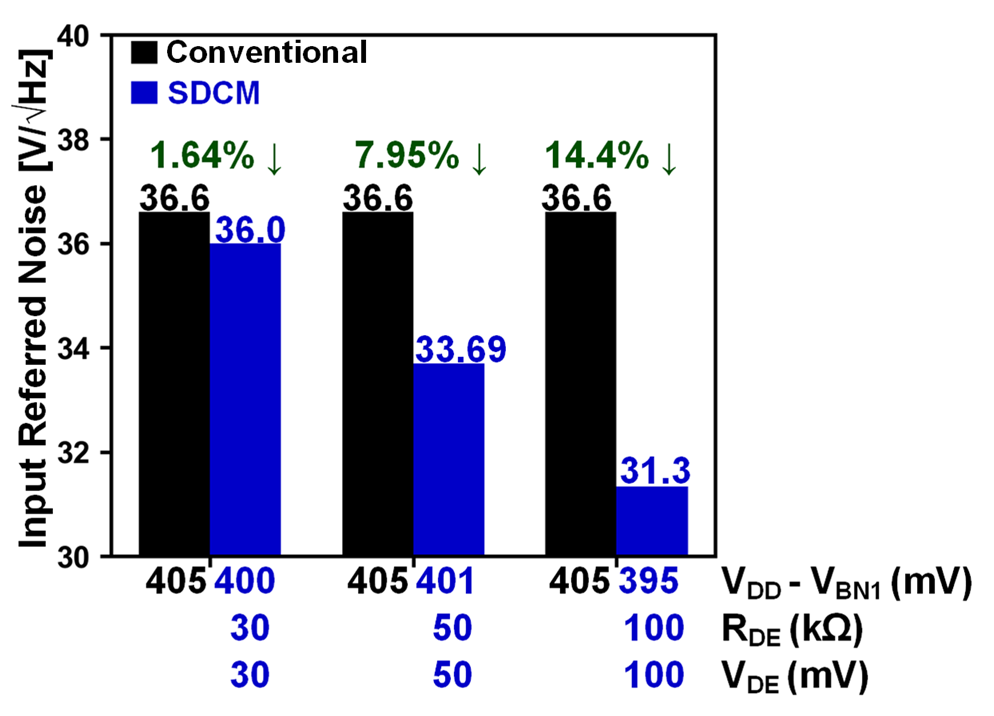}
    \vspace{-8mm}
    \caption{Comparison of the input referred noise performance between conventional current mirror and \ac{SDCM}, under the same voltage headroom near half-$V_\text{DD}$.}\label{fig:same voltage headroom}
    \end{center}
    \vspace{-5mm}
\end{figure}

\begin{figure}[t]
    \begin{center}
    \includegraphics[width=\columnwidth]{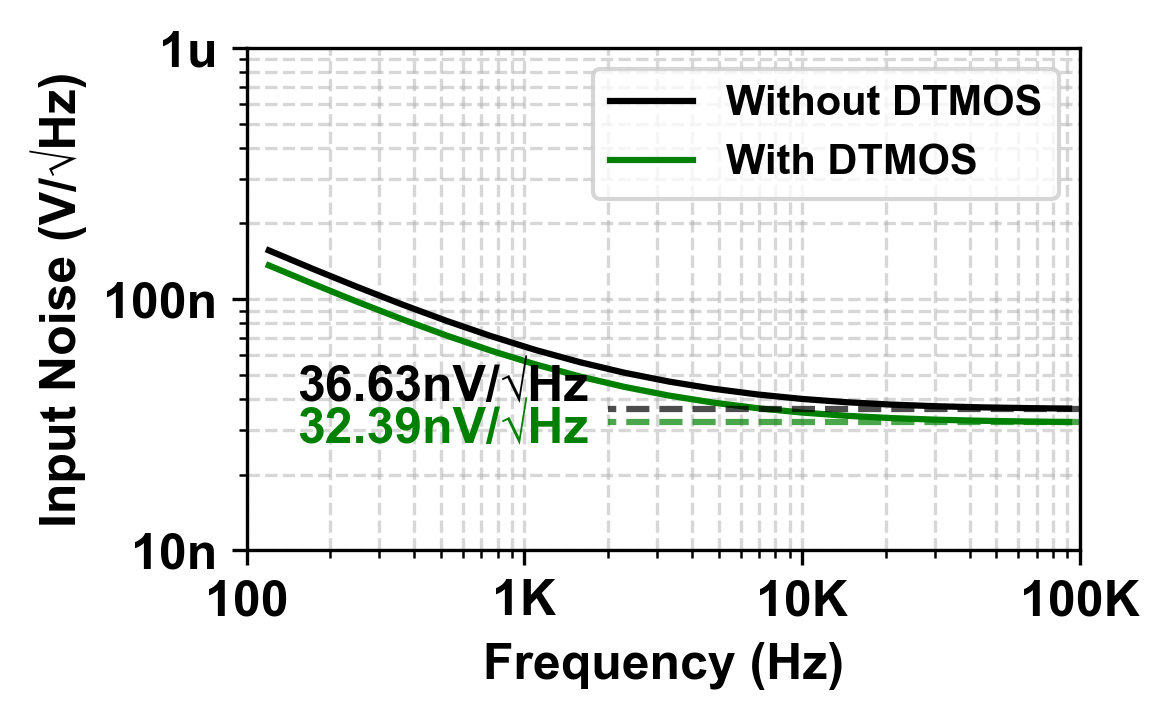}
    \vspace{-8mm}
    \caption{Input referred noise performance with/without the \ac{DTMOS} scheme.}\label{fig:DTMOS_noise}
    \end{center}
    \vspace{-5mm}
\end{figure}

\begin{figure}[t]
    \begin{center}
    \includegraphics[width=\columnwidth]{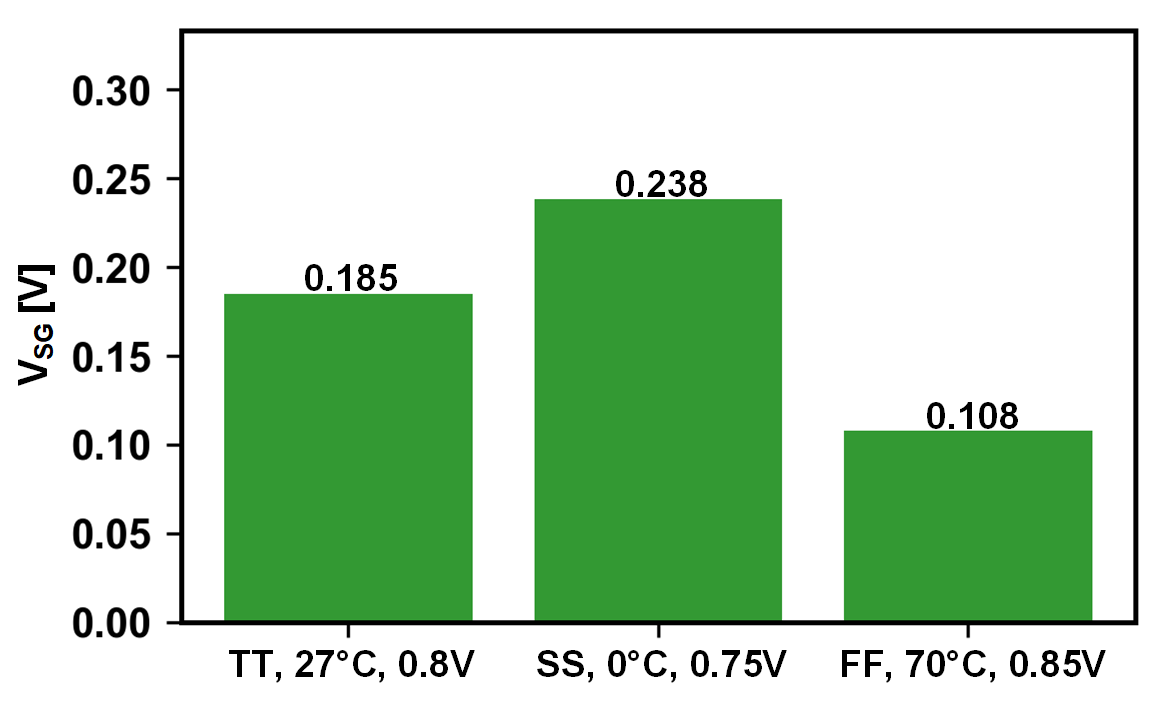}
    \vspace{-8mm}
    \caption{$V_\text{SG}$ of $M_{1}$ across PVT in the \ac{DTMOS} scheme .}\label{fig:Vsg_M1}
    \end{center}
    \vspace{-5mm}
\end{figure}

As shown in Fig.~\ref{fig:SDCM}(b) and (c), the \ac{SDCM} achieves noise reduction by increasing $R_\text{DE}$. However, a large $R_\text{DE}$ not only suppresses output noise but also elevates the source nodes of the transistors as well, leading to a degraded voltage headroom which is a similar problem of the conventional current mirror. Since excessively large $R_\text{DE}$ may drive $M_{3}$ out of saturation and into the linear region, $R_\text{DE}$ cannot be arbitrarily increased and must be carefully optimized.

Notably, Eq.~(\ref{gm}) indicates that $V_\text{GS}$ is inversely proportional to the transistor width for a fixed $I_\text{D}$. This property can be exploited as a design strategy to lower the diode-connected voltage $V_\text{DIO}$ at $M_{3m}$ and thereby recover the degraded headroom. As shown in Fig.~\ref{fig:SDCM}(d), (e), increasing the width has a negligible impact on thermal noise, demonstrating that a larger width is an effective strategy to recover voltage headroom on the $M_{3m}$ side from 450\,mV to 610\,mV without compromising noise performance. This is possible because Eq.~(\ref{eq:output_noise_Fi2.(b)}) shows that the output noise \ac{PSD} of the \ac{SDCM} is no longer proportional to $g_{m3}$, unlike the conventional current mirror described in Eq.~(\ref{eq:output_noise_Fi2.(a)}). In addition, the larger active area of the wider transistor in the \ac{SDCM} helps reduce flicker noise. However, the additional voltage penalty $V_\text{DE}$ introduced by the degeneration resistor $R_\text{DE}$ on the mirrored side $M_3$ remains unchanged regardless of transistor size, revealing a trade-off relationship between noise and headroom. Therefore, the designer must carefully optimize the transistor width and $R_\text{DE}$ to achieve both low output noise and sufficient voltage headroom without significantly consuming the available voltage budget on the mirrored side. 

By comparing Fig.~\ref{fig:LLCM}(b), (c) with Fig.~\ref{fig:SDCM}(d), (e), it can be observed that when both the conventional current mirror and \ac{SDCM} achieve a similar output noise level of approximately 230\,nV/$\surd$Hz, the \ac{SDCM} configured with a 100\,k$\Omega$ of $R_\text{DE}$ and a 20\,$\mu$m wide device width provides an additional 300\,mV of voltage headroom compared to the conventional current mirror. However, this improvement comes at the cost of an extra voltage penalty of approximately 100\,mV introduced by $V_\text{DE}$ on the mirrored side due to the degeneration resistor $R_\text{DE}$.

\subsection{Integration and Performance Comparison in IA}

To evaluate the practical impact of current mirror implementations within the \ac{IA}, we developed two versions of \ac{IA}, employing the conventional current mirror and the \ac{SDCM} integrated into the \ac{TC} stage. Fig.~\ref{fig:half_circuit} illustrates both versions of the \ac{TC} stage where differential mode half-circuit equivalents are drawn for simplicity, but the actual circuits are implemented in a differential architecture (see Fig.~\ref{fig:schematic_of_complete_IA}). We swept the length of the conventional current mirror in the first version and the width of the \ac{SDCM} in the second version.

Note that although the \ac{SDCM} approach has been widely used for noise reduction in current mirrors \cite{sansen2006essential, Yang2023neural-jssc, JSSC22Choi}, limited discussions have been conducted in prior research on optimizing both noise reduction and voltage headroom. Our work aims to address the lack of discussions on the trade-off between noise and voltage headroom of the \ac{SDCM}, when it is deployed in the \ac{TC} stage of the \ac{IA}.

As shown in Fig.~\ref{fig:half_circuit}, when both versions of the \ac{IA} achieve a similar noise level approximately 33\,nV/$\surd$Hz, the design with \ac{SDCM} improves the voltage headroom at the $M_{3m}$ branch by 145\,mV when $R_\text{DE}=\text{50\,k}\Omega$ was used. Fig.~\ref{fig:same voltage headroom} shows another comparison in which $R_\text{DE}$ is swept while ensuring a half $V_\text{DD}$ of voltage headroom at the $M_{3m}$ branch, $V_\text{DD}-V_\text{BN1}\approx\text{400\,mV}$ for a fair comparison, by optimizing the transistor dimensions. Under this condition, the \ac{SDCM} achieves lower noise performance for larger $R_\text{DE}$ values, reducing the input referred noise from 1.64\% to 14.4\% when $R_\text{DE}$ increases from $\text{30\,k}\Omega$ to $\text{100\,k}\Omega$. However, this benefit comes with an increase in the voltage drop $R_\text{DE}$ in the main signal path of the \ac{TC} stage, from 30\,mV at $R_\text{DE}=\text{30\,k}\Omega$ to 100\,mV at $R_\text{DE}=\text{100\,k}\Omega$.

In this work, we adopted $R_\text{DE}=\text{50\,k}\Omega$ in the design of \ac{IA} considering a negligible $V_\text{DE}=\text{50\,mV}$ penalty at the \ac{TC} stage but a 145\,mV improved voltage headroom at the $M_{3m}$ branch, achieving a 7.95\% higher noise efficiency under the same power consumption.

\section{dynamic threshold mos}\label{sec:dtmos}

\begin{figure*}[t]
    \begin{center}
    \includegraphics[width=\textwidth]{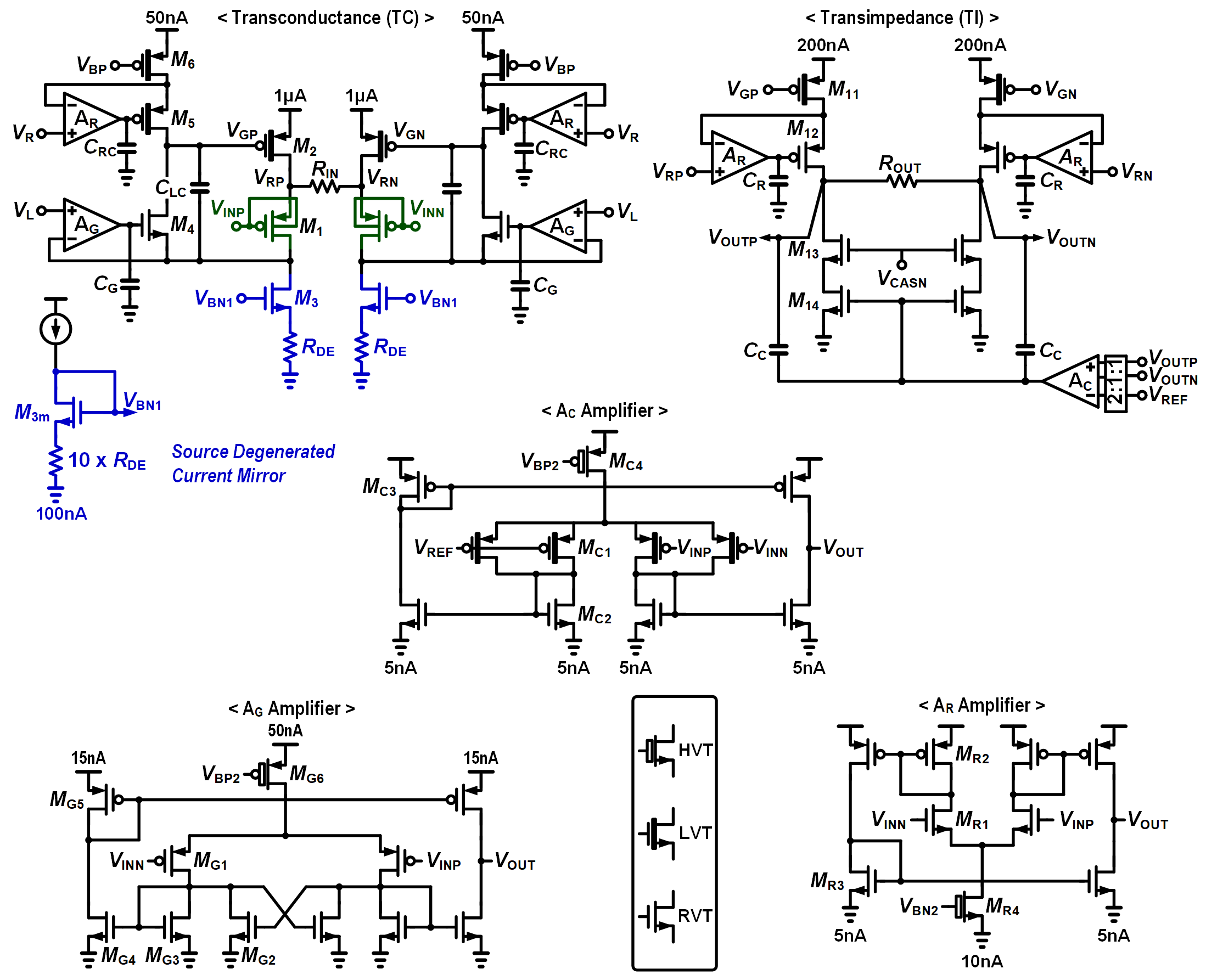}
    \caption{Schematic of the proposed \ac{IA}.}\label{fig:schematic_of_complete_IA}
    \end{center}
    \vspace{-5mm}
\end{figure*}

In conventional CMOS amplifier designs, the bulk terminal is typically connected to the source to eliminate the body effect by forcing $V_\text{SB} = 0$, to stabilize the variation of the threshold voltage $V_\text{TH}$. However, this configuration removes the bulk transconductance $g_{mb}$ and thus does not fully utilize the opportunity to exploit the potential knob to maximize the $g_{m}$ of the input transistor.

In this work, we connect the bulk terminal of the input transistor $M_1$ to its gate as shown in Fig.~\ref{fig:schematic_of_complete_IA}, forming a \ac{DTMOS}~\cite{Assaderaghi1997TED}. This technique introduces a positive $V_\text{SB}$ equal to the source-gate voltage $V_\text{SG}$, which dynamically modulates the threshold voltage $V_\text{TH}$ and enhances the overall transconductance through body effect. The governing equations are as follows.
\begin{equation} \label{eq:threshold_voltage}
V_{\text{TH}} = V_{\text{TH}0} + \lambda \left( \sqrt{2\Phi_\text{F} + V_{\text{SB}}} - \sqrt{2\Phi_\text{F}} \right)
\end{equation}

\begin{equation} \label{eq:gmb1}
g_{mb} = \frac{\partial I_\text{D}}{\partial V_\text{BS}}
= \mu_n C_{\text{ox}} \frac{W}{L} 
\left( V_\text{GS} - V_\text{TH} \right) 
\left( - \frac{\partial V_\text{TH}}{\partial V_\text{BS}} \right)
\end{equation}
\begin{equation} \label{eq:Vth_Vbs}
\frac{\partial V_\text{TH}}{\partial V_\text{BS}}
= - \frac{\partial V_\text{TH}}{\partial V_\text{SB}}
= -\frac{\lambda}{2} \left( 2\Phi_\text{F} + V_\text{SB} \right)^{-1/2}
\end{equation}
\begin{equation} \label{eq:gmb}
g_{mb}
= \frac{\partial I_\text{D}}{\partial V_\text{SB}}
= g_{m} \frac{\lambda}{2\sqrt{2\phi_\text{F} + V_{\text{SB}}}}
\end{equation}
where $\lambda$ denotes the body-effect coefficient and $2\Phi_\text{F}$ is the inversion layer potential. Since the transconductance $g_{m}$ of $M_{1}$, as indicated by Eq.~(\ref{eq:noise_IA_baseline}), is inversely proportional to the input referred noise, the additional body transconductance $g_{mb}$ in Eq.~(\ref{eq:gmb}) introduced by the \ac{DTMOS} boosts the effective transconductance of $M_1$ from $g_{m1}$ to $g_{m1}+g_{mb1}$ thereby improving the noise performance. Fig.~\ref{fig:DTMOS_noise} compares the noise performance of the \ac{IA} when the \ac{TC} stage employs the \ac{DTMOS} scheme over the baseline case. By dynamically enhancing effective transconductance through additional \(g_{mb}\), the \ac{DTMOS} configuration achieves a remarkable reduction in thermal noise. At 100\,kHz, the input referred noise decreases from \text{36.63}\,nV/$\surd{\text{Hz}}$ to \text{32.39}\,nV/$\surd{\text{Hz}}$, corresponding to an improvement of approximately 11.5\%. Note that this noise benefit is achieved without increasing power consumption, as the \ac{DTMOS} leverages body effect rather than higher bias currents to enhance the effective transconductance.

However, a potential issue of the \ac{DTMOS} technique is the risk of forward biasing of the source-body junction. Since the bulk terminal is connected to the gate, a large positive $V_\text{SB}$ may forward bias the source-body junction and activate the $pn$ diode, which can lead to excessive leakage current or even transistor malfunction. Therefore, the maximum allowable $V_\text{SG}$ must be carefully constrained to avoid forward conduction while maintaining the intended noise advantage.

To validate the safe operating region of the proposed DTMOS configuration, a comprehensive PVT analysis was performed on the input PMOS $M_{1}$, as illustrated in Fig.~\ref{fig:Vsg_M1}. With the body and gate shorted, $V_\text{SB}$ is equal to $V_\text{SG}$, and thus analyzing $V_\text{SG}$ allows us to directly determine the forward-biasing condition of the body diode. Since the input transistor $M_1$ is typically optimized to operate in the subthreshold region to maximize $g_m$, $V_\text{SG}$ could be kept lower than or close to $V_\text{TH}$. In our design, the worst-case simulated $V_\text{SG}$ values are 185\,mV, 238\,mV, and 108\,mV, all comfortably below 250\,mV. Since parasitic diodes cannot be activated with such a negligible amount of forward bias \cite{Chatterjee2005JSSC}, it is confirmed that the proposed \ac{DTMOS} structure operates safely under the PVT corners while preserving the intended $g_{m}$ benefits.

\begin{figure*}[t]
    \begin{center}
    \includegraphics[width=\textwidth]{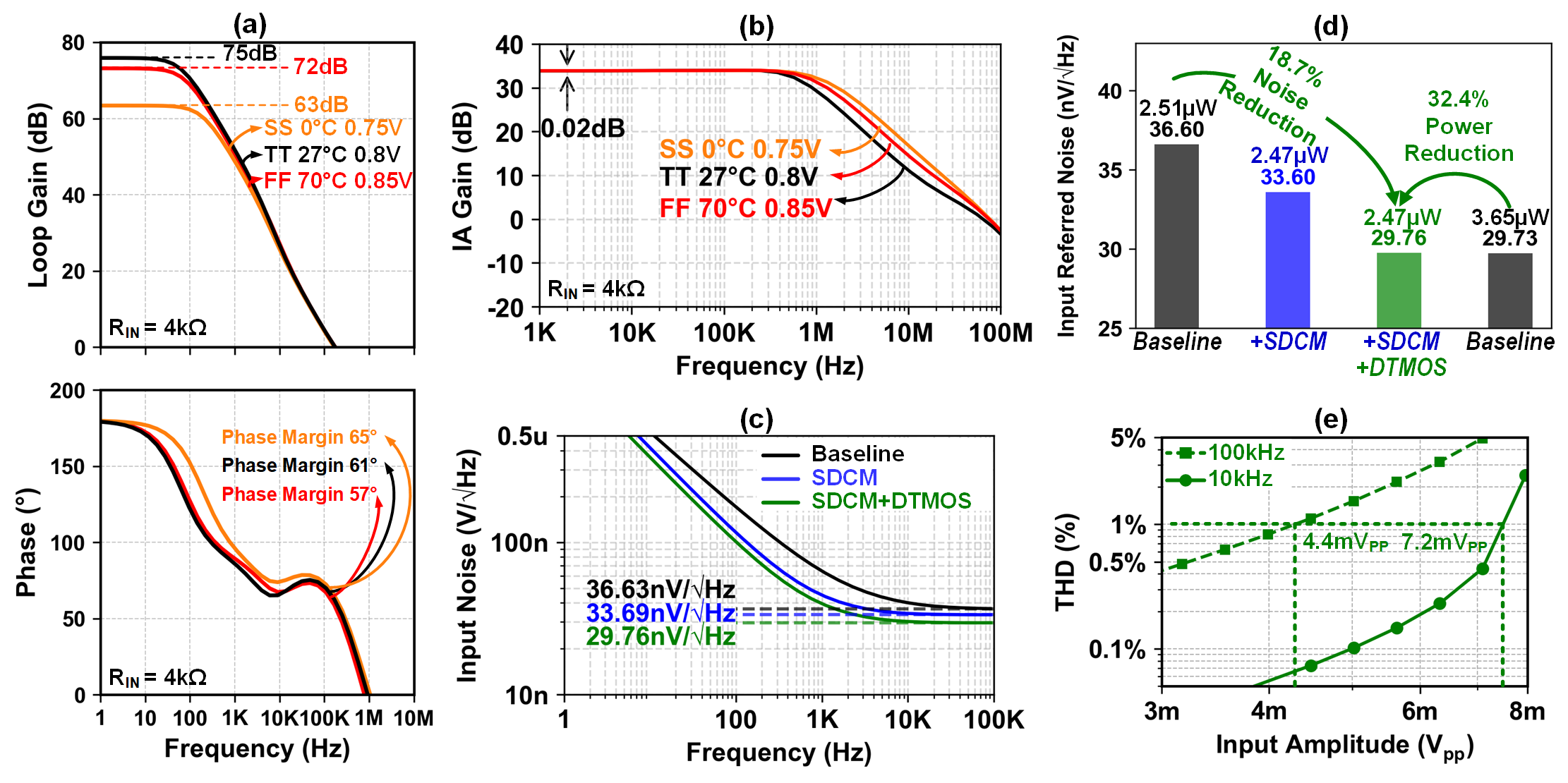}
    \caption{Simulated (a) differential mode loop response of the \ac{FVF}, (b) overall gain of the \ac{IA}, (c) input referred noise, and (d) noise comparison. In (d), $V_\text{DD}-V_\text{BN1}$ in Fig.~\ref{fig:schematic_of_complete_IA}, are 405\,mV, 408\,mV, 408\,mV, 387\,mV from the leftmost to rightmost bar, respectively, and (e) total harmonic distortion (\ac{THD}) performance versus input amplitude at \text{10}\,kHz and \text{100}\,kHz.}\label{fig:simulation_results}
    \end{center}
    \vspace{-5mm}
\end{figure*}

\begin{figure}[t]
    \begin{center}
    \includegraphics[width=\columnwidth]{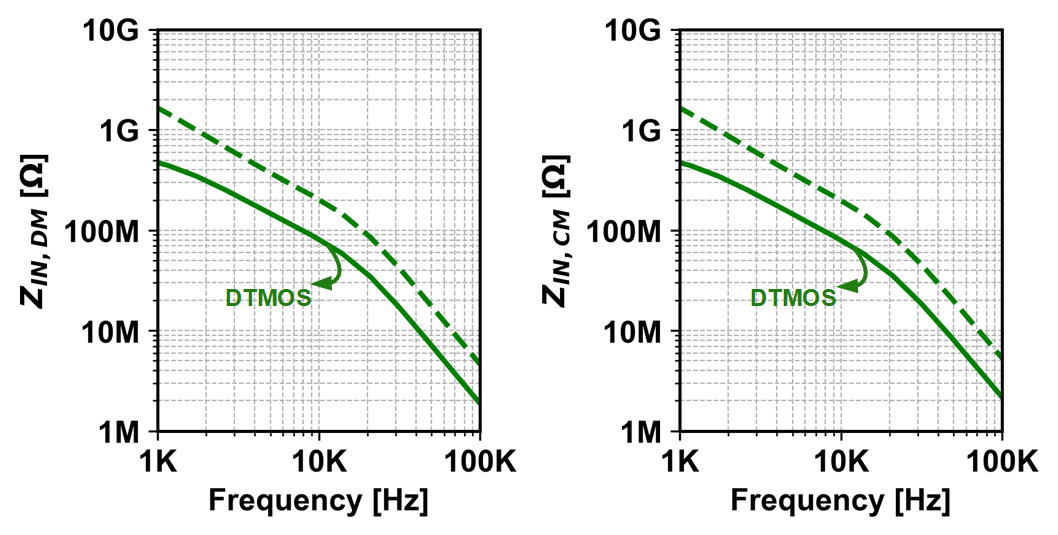}
    \caption{Input impedance of the \ac{IA} with/without the \ac{DTMOS} scheme.}\label{fig:RIN}
    \end{center}
    \vspace{-5mm}
\end{figure}

\section{simulation results}\label{sec:proposed IA}

Fig.~\ref{fig:schematic_of_complete_IA} shows the complete transistor-level schematic of the proposed \ac{IA} using the 28\,nm CMOS process, which incorporates the \ac{SDCM} and the \ac{DTMOS} schemes into the \ac{TC} stage for low-noise, low-power operation. The detailed device parameters are summarized in Table~\ref{tab:full_params_one_table}. A similar design strategy, as used in prior works \cite{Kim2020BioZ, Kim2023BioZ-ISCAS}, was adopted to optimize the \ac{TC} stage, including a sufficiently large width in $M_1$ and assigning most of the current to the main branch (1\,$\mu$A), to maximize $g_{m1}$ thereby suppressing input referred noise (see Eq.~(\ref{eq:noise_IA_baseline})). The lead compensation scheme ($C_\text{LC}$) \cite{Kim2020BioZ} was adopted for a power-bandwidth-efficient loop stabilization. We increased the width of the $M_2$ transistor in \ac{TC} stage to reduce its $V_\text{SG}$ drop, to keep the voltage level of the $V_\text{GP}$ node high. As such, a sufficient amount of voltage headroom could be given to the $M_4$ transistor, resulting in the proper operation of the \ac{IA} across PVT corners (see Fig.~\ref{fig:simulation_results}). Two auxiliary amplifiers in the \ac{TC} stage and one in the \ac{TI} stage were used for proper biasing: the drain regulation amplifier $A_\text{R}$, the gain-boosting amplifier $A_{G}$, and the common-mode feedback amplifier $A_{C}$ \cite{Kim2023BioZ-ISCAS, Kim2020BioZ} where each current consumption was kept minimized to boost power efficiency. Within the TC stage, $A_{\text{R}}$ not only ensures a proper distribution of $V_{\text{DS}}$ of $M_{5}$ and $M_{6}$ but also acts as a gain-boosting amplifier, effectively increasing the output impedance and improving current mirroring accuracy. The $A_{\text{G}}$ amplifier employs a positive feedback topology to significantly enhance its effective gain\cite{Kim2023BioZ-ISCAS}. In the TI stage, an additional drain-regulation loop based on $A_{\text{R}}$ is deployed to maintain accurate current mirroring from the TC stage. The $A_{\text{C}}$ amplifier operates as a common-mode feedback circuit in the TI stage, continuously monitoring $V_{\text{OUTP}}$ and $V_{\text{OUTN}}$ to maintain a stable DC bias point.



Fig.~\ref{fig:simulation_results} presents the key simulation results of the proposed \ac{IA}, including the \ac{FVF} loop response, input-referred noise, overall gain, and a comparative noise and power analysis with and without the proposed techniques. Using a gain-boosted cascode current source ($M_5$, $M_6$, and $A_\text{R}$) together with a gain-boosted common-gate amplifier ($M_{4}$ and $A_\text{G}$), the \ac{IA} achieves 75\,dB differential-mode DC loop gain in the \ac{FVF} loop even with a small $R_\text{IN}$ of 4\,k$\Omega$. This high loop gain is maintained across \ac{PVT} corners. As a result, Fig.~\ref{fig:simulation_results}(b) shows a robust 34\,dB closed-loop AC gain across \ac{PVT} corners, with less than 0.02\,dB variation, while the nominal case achieves a bandwidth of 1.44\,MHz.
As illustrated in Fig.~\ref{fig:simulation_results}(c), applying the \ac{SDCM} and \ac{DTMOS} techniques reduces the input-referred noise from \text{36.63}\,nV/$\surd\text{Hz}$ in the baseline design to \text{33.69}\,nV/$\surd\text{Hz}$ and \text{29.76}\,nV/$\surd\text{Hz}$, corresponding to \text{7.95\%} and \text{11.7\%} reductions, achieving an overall improvement of \text{18.7\%}. Fig.~\ref{fig:simulation_results}(d) compares the baseline and proposed \ac{IA} under the same $V_{\mathrm{DD}}-V_{\mathrm{BN1}}$ biasing condition, showing that the proposed design achieves equivalent noise performance while consuming 32.4\% less power compared to the baseline \ac{IA} under iso-noise condition, demonstrating superior noise efficiency. This amount of power reduction makes a good agreement with the theoretical estimation, i.e., $(1-0.187)^2=1-0.339$, corresponding to a 33.9\% reduction that is close to our result of 32.4\%. This agreement follows from the fact that the input referred RMS noise of low-power amplifiers is inversely proportional to the square root of the current, e.g., the \ac{NEF} \cite{Steyaert1987JSSC-NEF, Harrison2003JSSC}. Fig.~\ref{fig:simulation_results}(e) illustrates \ac{THD} performance (up to \text{10\textsuperscript{th}} harmonic) at both \text{10}\,kHz and \text{100}\,kHz using the PSS simulation in the Spectre environment, confirming that the circuit maintains \text{1\%} \ac{THD} with 7.2\,mV\textsubscript{PP} and 4.4\,mV\textsubscript{PP} input amplitudes, respectively.

While the proposed \ac{DTMOS} technique effectively improves transconductance and reduces input-referred noise, it introduces a trade-off in the input impedance characteristics. 
As shown in Fig.~\ref{fig:RIN}, connecting the bulk terminal to the gate effectively adds an additional parasitic capacitance at the input node, which lowers both the differential-mode and common-mode input impedances across the entire frequency range. Although the achieved input impedance remains sufficient ($\approx$7\,M$\Omega$ at 50\,kHz) for most wet type, four-electrode \ac{BioZ} sensing scenarios, where its typical value ranges from 500\,$\Omega$-2\,k$\Omega$ depending on measurement frequency and contact status \cite{Ha2019JSSC}, this reduction could affect signal integrity in ultra-high-impedance electrodes such as $<$1\,cm\textsuperscript{2} dry type electrodes \cite{Pan2022BioZ}. Future work may focus on optimizing the \ac{DTMOS} scheme or introducing impedance boosting techniques \cite{Pan2022BioZ, Zhong2022adc-a-sscc} to mitigate the adverse impact on the input impedance without compromising the noise benefit.

The simulated performance of the proposed \ac{IA} is summarized in Table~\ref{table:Proposed_IA} and compared with other recent state-of-the-art \ac{IA} designs. Our design with $R_\text{IN}=4\,\text{k}\Omega$ achieves a 21\% lower power consumption, 26\% lower input referred noise, 3.9x higher bandwidth than \cite{Kim2023BioZ-ISCAS}, while achieving a similar input linear range even under the 20\% lower supply voltage. Compared to \cite{Yan2013TBioCAS, Kim2017BioZ-ESSCIRC}, our design with $R_\text{IN}=20\,\text{k}\Omega$ achieves superior noise and bandwidth performance while maintaining a comparable input range. Notably, this input range is obtained even at a supply voltage of 0.8\,V despite the reduced headroom, which is significantly lower than 1.8\,V used in \cite{Yan2013TBioCAS, Kim2017BioZ-ESSCIRC}.

\begin{table*}[t]
    \centering
    \caption{Performance Comparison Table - Low-Power \ac{IA} ($<$10\,$\mu$W) for \ac{BioZ} Sensors}
    \label{table:Proposed_IA}
    \begin{tabular}{||c||c|c|c|c|c|c|c||}
        \hline
        \textbf{Low-Power IA} &
        \begin{tabular}[c]{@{}c@{}}
        TBioCAS'13 \\ \cite{Yan2013TBioCAS} \end{tabular} &
        \begin{tabular}[c]{@{}c@{}}
        ESSCIRC'17 \\ \cite{Kim2017BioZ-ESSCIRC} \end{tabular} &
        \begin{tabular}[c]{@{}c@{}}
        JSSC'20 \\ \cite{Kim2020BioZ}
        \end{tabular} &
        \begin{tabular}[c]{@{}c@{}}
        TBioCAS'22 \\ \cite{Rezaee-Dehsorkh2022BioZ}
        \end{tabular} &
        \begin{tabular}[c]{@{}c@{}}
        ISCAS'23 \\ \cite{Kim2023BioZ-ISCAS}
        \end{tabular} &
        \begin{tabular}[c]{@{}c@{}}\textbf{This Work} \\ ($R_\text{IN}=4\,\text{k}\Omega$) \end{tabular} &
        \begin{tabular}[c]{@{}c@{}}\textbf{This Work} \\ ($R_\text{IN}=20\,\text{k}\Omega$) \end{tabular} \\
        \hline
        Process (nm) & 180 & 180 & 65 & 180 & 65 & \multicolumn{2}{c||}{\textbf{28}} \\ \hline
        Supply (V) & 1.8 & 1.8 & 0.5 & 1.0 & 1.0 & \multicolumn{2}{c||}{\textbf{0.8}} \\ \hline
        Gain (dB) & 30 & 36 & 31 & 51 & 40 &
        \textbf{34} &
        \textbf{20} \\
        \hline
        Power ($\mu$W) & 2.16 & 2.12 & 3.95 & 0.7 & 3.11 & \multicolumn{2}{c||}{\textbf{2.47}} \\ \hline
        Noise (nV/$\surd$Hz) &
        149 &
        115 &
        45 &
        300{\color{Maroon}\textsuperscript{A}} &
        40 &
        \textbf{29.76} &
        \textbf{46.18} \\
        \hline
        Bandwidth (Hz) & 20\,k & 20\,k & 408\,k & 3\,k & 369\,k &
        \textbf{1.44\,M} &
        \textbf{4.67\,M} \\
        \hline
        Input Range (mV\textsubscript{PP}) &
        44 (20\,kHz) &
        52 (20\,kHz) &
        1.6 (20\,kHz) &
        10 &
        7.26 (10\,kHz) &
        \begin{tabular}[c]{@{}c@{}}
        \textbf{7.2 (10\,kHz)} \\
        \textbf{4.4 (100\,kHz)}
        \end{tabular} &
        \begin{tabular}[c]{@{}c@{}}
        \textbf{40 (10\,kHz)} \\
        \textbf{32 (100\,kHz)}
        \end{tabular} \\
        \hline
        \multicolumn{8}{|l|}{{\color{Maroon}\textsuperscript{A}}Estimated from figure} \\
        \hline
    \end{tabular}
\end{table*}

\begin{table}[t]
  \centering
  \caption{Design parameters of proposed \ac{IA}.}
  \resizebox{0.95\linewidth}{!}{
  \begin{tabular}{|c|c|c|c|}
    \hline
    \text{TC stage} & W/L ($\mu$m) & A\textsubscript{R} Amplifier & W/L ($\mu$m) \\ \hline
    $M_{1}$   & 500/0.25  & $M_\text{R1}$ & 0.25/1 \\
    $M_{2}$   & 30/0.5    & $M_\text{R2}$ & 1/1 \\
    $M_{3m}$  & 2/1       & $M_\text{R3}$ & 0.25/0.25 \\
    $M_{3}$   & 20/1      & $M_\text{R4}$ & 0.25/1 \\
    $M_{4}$   & 0.25/1    & $C_\text{RC}$ & 10\,fF \\ \cline{3-4}
    $M_{5}$   & 1/1       & A\textsubscript{G} Amplifier & W/L ($\mu$m) \\ \cline{3-4}
    $M_{6}$   & 1/1       & $M_{G1}$ & 0.25/0.25 \\
    $V_\text{R}$   & \multicolumn{1}{c|}{100\,mV} & $M_\text{G2}$ & 0.8/1 \\
    $V_\text{L}$   & \multicolumn{1}{c|}{250\,mV} & $M_\text{G3}$ & 1/1 \\
    $C_\text{LC}$  & \multicolumn{1}{c|}{750\,fF} & $M_\text{G4}$ & 1/1 \\
    $R_\text{DE}$  & \multicolumn{1}{c|}{50\,k$\Omega$} & $M_\text{G5}$ & 0.25/0.25 \\
    $R_\text{IN}$  & \multicolumn{1}{c|}{4\,k$\Omega$ or 20\,k$\Omega$}  & $M_\text{G6}$ & 0.3/1 \\
              &                  & $C_\text{G}$ & 5\,pF \\ \hline\hline

    \text{TI stage} & W/L ($\mu$m) & A\textsubscript{C} Amplifier & W/L ($\mu$m) \\ \hline
    $M_{11}$ & 6/0.5   & $M_\text{C1}$ & 1/1 \\
    $M_{12}$ & 1/0.25  & $M_\text{C2}$ & 0.25/0.25 \\
    $M_{13}$ & 0.5/0.5 & $M_\text{C3}$ & 0.25/0.25 \\
    $M_{14}$ & 0.25/4  & $M_\text{C4}$ & 0.25/1 \\
    $C_\text{C}$  & 5\,fF   & $V_\text{REF}$ & 400\,mV \\
    $C_\text{R}$  & 20\,fF  &          & \\
    $R_\text{OUT}$& $250\times R_\text{IN}$ & & \\
    \hline
  \end{tabular}}
  \label{tab:full_params_one_table}
\end{table}

\section{conclusion}\label{sec:conclusion}

We have presented a low-power, low-noise \ac{IA} for \ac{BioZ} sensing applications, which exploits \ac{SDCM} and \ac{DTMOS} techniques. Simulated in a 28\,nm CMOS process, it consumes 2.5\,$\mu$W power dissipation with a power supply of 0.8\,V, while demonstrating an input-referred noise of 30\,$\text{nV}/\surd{\text{Hz}}$ within a 1.44\,MHz bandwidth at $R_{\text{IN}}=4\,\text{k}\Omega$ and $R_{\text{DE}}=50\,\text{k}\Omega$. Discussions on the noise-headroom trade-off in \ac{SDCM}, and the noise-input impedance trade-off in \ac{DTMOS} are provided. By combining the \ac{SDCM} and \ac{DTMOS} techniques, the input-referred noise is reduced by 18.7\%, which is translated to a 32.4\% power saving under the same noise performance compared to the baseline design.

By providing a low-power, low-noise core amplifier structure for \ac{BioZ} sensors, our design offers a fundamental building block that can be seamlessly integrated into advanced voltage readout techniques recently published in the literature, such as baseline cancellation \cite{Ha2019JSSC, Zhang2023BioZ-JSSC}, differential difference amplifiers \cite{Cheon2024bioz-isscc}, and $\Delta\Sigma$ modulators \cite{Zhang2023BioZ-JSSC}. The complete design parameters have been open-sourced to promote reproducibility and support future research developments.

{\appendix

This appendix presents the derivation of the output-referred noise in the source-degenerated current mirror (SDCM), as described in Eq.~(\ref{eq:output_noise_Fi2.(b)}).

\begin{figure}[h]
    \centering
    \includegraphics[width=0.3\columnwidth]{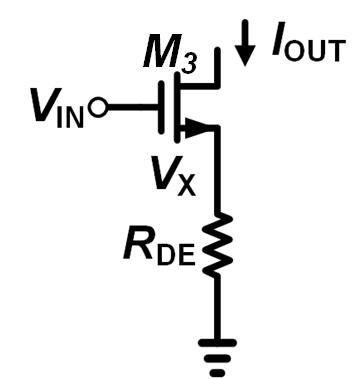}
    \caption{Common-source stage with source degeneration.}
    \label{fig:appendix_sdcmeq1}
\end{figure}

The output noise current consists of two primary contributions: 1) the thermal noise of the degeneration resistor $R_{\text{DE}}$, and 2) the thermal noise of the mirroring transistor $M_3$. In this analysis, we use the superposition principle of the noise \ac{PSD} after calculating each noise contribution, since two noise sources are uncorrelated. Note that noise currents cause small-signal voltage fluctuations $v_x$ that force the transistor to behave as a transconductor, injecting a small signal current $g_{m3} v_{gs3}$ back into the $v_x$ node, thereby creating feedback. Due to this local negative feedback introduced by $R_{\text{DE}}$ \cite{Kim2023filt-casm}, the impact of the noise coming from the mirroring transistor $M_3$ is attenuated by a factor related to the loop gain $g_{m3} R_{\text{DE}}$, similar to the case of noise contribution of cascode devices (see Chapter 7.4.4 in \cite{razavi2001design}).

\begin{figure}[h]
    \begin{center}
    \includegraphics[width=0.6\columnwidth]{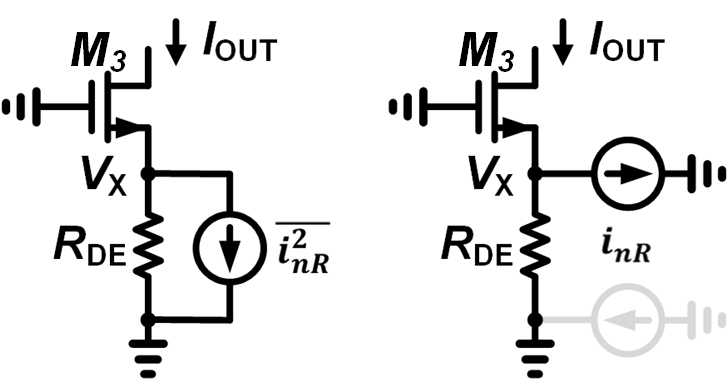}
    \caption{Current noise model for resistor $R_\text{DE}$ in the feedback loop.}\label{fig:current noise model of RDE}
    \end{center}
\end{figure}

First, let us analyze the thermal noise current generated by $R_{\text{DE}}$. We use the correlated decomposition model of current noise (see Chapter 7.6 in \cite{razavi2001design}). As shown in Fig.~\ref{fig:current noise model of RDE}, the noise current can be decomposed into two correlated current sources where the noise current connected to GND in both terminals can be omitted since its noise current flows directly to GND, leaving the circuit unaffected. The resulting noise current contributed by $R_\text{DE}$, e.g., $i_{n,\text{out}}|_{R_\text{DE}}$, on the drain side of the transistor can be expressed as the following, when we neglect channel-length modulation:
\begin{equation}
    i_{n,\text{out}}|_{R_\text{DE}} = \frac{v_x}{R_{\text{DE}}} + i_{nR}.
\end{equation}

Since this current must be equal to the small-signal current from the transistor:
\begin{equation}
    i_{n,\text{out}}|_{R_\text{DE}} = -g_{m3} v_x.
\end{equation}

Solving for $v_x$ gives:
\begin{equation}
    v_x = -\frac{R_{\text{DE}}}{1 + g_{m3} R_{\text{DE}}} i_{nR},
\end{equation}
which leads to:
\begin{equation}
    i_{n,\text{out}|R_{\text{DE}}} = \frac{g_{m3} R_{\text{DE}}}{1 + g_{m3} R_{\text{DE}}} i_{nR}.
\end{equation}

\begin{figure}[h]
    \begin{center}
    \includegraphics[width=0.6\columnwidth]{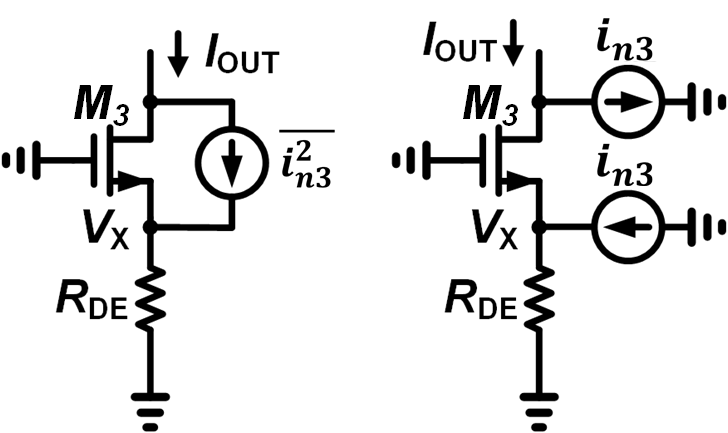}
    \caption{Current noise model for $M_\text{3}$ in the feedback loop.}\label{fig:current noise model of M3}
    \end{center}
\end{figure}

Now, consider the noise contributed by $M_3$, e.g., $i_{n,\text{out}}|_{M_3}$, as shown in Fig.~\ref{fig:current noise model of M3}:
\begin{equation}
    i_{n,\text{out}}|_{M_3} = \frac{v_x}{R_{\text{DE}}} = -g_{m3} v_x + i_{n3}.
\end{equation}

Solving again gives:
\begin{equation}
    v_x = \frac{R_{\text{DE}}}{1 + g_{m3} R_{\text{DE}}} i_{n3},
\end{equation}
and:
\begin{equation}
    i_{n,\text{out}|M_3} = \frac{1}{1 + g_{m3} R_{\text{DE}}} i_{n3}.
\end{equation}

Combining both components in the power domain, the total output current noise becomes:
\begin{align}
    \overline{i_{n,\text{out}}^2}
    &= \overline{i_{n,\text{out}}^2}|_{R_\text{DE}}
    + \overline{i_{n,\text{out}}^2}|_{M3} \nonumber\\
    &= \left( \frac{g_{m3} R_{\text{DE}}}{1 + g_{m3} R_{\text{DE}}} \right)^2 \overline{i_{nR}^2}
    + \left( \frac{1}{1 + g_{m3} R_{\text{DE}}} \right)^2 \overline{i_{n3}^2}.
\end{align}

Substituting the following:
\[
\overline{i_{nR}^2} = \frac{4kT}{R_{\text{DE}}}, \quad \overline{i_{n3}^2} = 4kT \gamma g_{m3},
\]
we obtain the total output voltage noise of \ac{SDCM} shown in Fig.~\ref{fig:SDCM} and Eq.~(\ref{eq:output_noise_Fi2.(b)}):
\begin{align}
    \overline{v_{n,\text{SDCM}}^2}
    &= \bigg( \frac{4kT\gamma g_{m3} + (g_{m3} R_\text{DE})^2 4kT/R_\text{DE}}{(1+g_{m3} R_\text{DE})^2} \bigg)R_\text{D}^2 \nonumber\\
    &\quad + \overline{v_\text{RD}^2}
\end{align}




\bibliographystyle{IEEEtran}
\bibliography{myref}

@INPROCEEDINGS{Yu2025Newcas,
  author={Xue, Yu and Kim, Kwantae},
  booktitle={2025 23rd IEEE Interregional NEWCAS Conference (NEWCAS)}, 
  title={{A 2.48$\,\mu\text{W}$ \text{33.13}$\,\text{nV}/\surd{\text{Hz}}$ Instrumentation Amplifier for Bio-Impedance Sensors with Source Degenerated Current Mirror and DTMOS Transistor}}, 
  year={2025},
  volume={},
  number={},
  pages={371-375},
  doi={10.1109/NewCAS64648.2025.11107006}}

@INPROCEEDINGS{Kim2017BioZ-ESSCIRC,
  author={Kim, Kwantae and Song, Kiseok and Bong, Kyeongryeol and Lee, Jaehyuk and Lee, Kwonjoon and Lee, Yongsu and Ha, Unsoo and Yoo, Hoi-Jun},
  booktitle={IEEE European Solid State Circuits Conference (ESSCIRC)}, 
  title={{A 24 $\mu$W 38.51 m$\Omega$rms Resolution Bio-Impedance Sensor with Dual Path Instrumentation Amplifier}}, 
  year={2017},
  volume={},
  number={},
  pages={223-226},
  doi={10.1109/ESSCIRC.2017.8094566}}

@ARTICLE{JSSC06Arnaud,
  author={Arnaud, A. and Fiorelli, R. and Galup-Montoro, C.},
  journal={IEEE Journal of Solid-State Circuits (JSSC)}, 
  title={{Nanowatt, Sub-nS OTAs, With Sub-10-mV Input Offset, Using Series-Parallel Current Mirrors}}, 
  year={2006},
  volume={41},
  number={9},
  pages={2009-2018},
  doi={10.1109/JSSC.2006.880606}}

@INPROCEEDINGS{Teng2014BioZ-Biocas,
  author={Teng, Yueh-Ching and Odame, Kofi M.},
  booktitle={IEEE Biomedical Circuits and Systems Conference (BioCAS)}, 
  title={{A 10 MHz 85 dB dynamic range instrumentation amplifier for electrical impedance tomography}}, 
  year={2014},
  volume={},
  number={},
  pages={632-635},
  doi={10.1109/BioCAS.2014.6981805}}

@ARTICLE{Kim2023BioZ-ISCAS,
  author={Kim, Kwantae and Shih-Chii Liu},
  journal={IEEE International Symposium on Circuits and Systems (ISCAS)},
  title={{A 3.11~$\mu$W 40 nV/$\surd$Hz Instrumentation Amplifier for Bio-Impedance Sensors Exploiting Positive-Feedback-Assisted Gain Boosting}}, 
  year={2023},
  volume={},
  number={},
  month={July .},
  pages={1-5},
  doi={10.1109/ISCAS46773.2023.10181417}
}

@ARTICLE{Kim2017EIT,
  author={Kim, Minseo and Jang, Jaeeun and Kim, Hyunki and Lee, Jihee and Lee, Jaehyuck and Lee, Jiwon and Lee, Kyoung-Rog and Kim, Kwantae and Lee, Yongsu and Lee, Kyuho Jason and Yoo, Hoi-Jun},
  journal={IEEE Journal of Solid-State Circuits (JSSC)}, 
  title={{A 1.4-m$\Omega$-Sensitivity 94-dB Dynamic-Range Electrical Impedance Tomography SoC and 48-Channel Hub-SoC for 3-D Lung Ventilation Monitoring System}}, 
  year={2017},
  month={Nov.},
  volume={52},
  number={11},
  pages={2829-2842},
  doi={10.1109/JSSC.2017.2753234}}

@ARTICLE{Kim2020BioZ,
  author={Kim, Kwantae and Kim, Ji-Hoon and Gweon, Surin and Kim, Minseo and Yoo, Hoi-Jun},
  journal={IEEE Journal of Solid-State Circuits (JSSC)}, 
  title={{A 0.5 V Sub-10 $\mu$W 15.28 m$\Omega$/$\surd$Hz Bio-Impedance Sensor IC with Sub-1$^{\circ}$ Phase Error}}, 
  year={2020},
  month={Aug.},
  volume={55},
  number={8},
  pages={2161-2173},
  doi={10.1109/JSSC.2020.2991511}}

@INPROCEEDINGS{Cheon2024bioz-isscc,
  author={Cheon, Song-I and Choi, Haidam and Yun, Gichan and Oh, Sein and Suh, Ji-Hoon and Ha, Sohmyung and Je, Minkyu},
  booktitle={IEEE International Solid-State Circuits Conference (ISSCC)}, 
  title={{A Two-Electrode Bio-Impedance Readout IC with Complex-Domain Noise-Correlated Baseline Cancellation Supporting Sinusoidal Excitation}}, 
  year={2024},
  volume={67},
  number={},
  pages={556-558},
  doi={10.1109/ISSCC49657.2024.10454399}}

@ARTICLE{Jiang2020eit-tcasii,
  author={Jiang, Dai and Wu, Yu and Demosthenous, Andreas},
  journal={IEEE Transactions on Circuits and Systems II: Express Briefs}, 
  title={{Hand Gesture Recognition Using Three-Dimensional Electrical Impedance Tomography}}, 
  year={2020},
  volume={67},
  number={9},
  pages={1554-1558},
  doi={10.1109/TCSII.2020.3006430}}

@ARTICLE{Ollmar2023glucose-sr,
  author={Ollmar, Stig and Fernandez Schrunder, Alejandro and Birgersson, Ulrik and Kristoffersson, Tomas and Rusu, Ana and Thorsson, Elina and Hedenqvist, Patricia and Manell, Elin and Rydén, Anneli and Jensen‑Waern, Marianne and Rodriguez, Saul},
  journal={Scientific Reports}, 
  title={{A battery‑less implantable glucose sensor based on electrical impedance spectroscopy}}, 
  year={2023},
  volume={13},
  number={1},
  pages={18122}}

@ARTICLE{Pan2024bioz-jssc,
  author={Pan, Qinjing and Luo, Qi and Qu, Tianxiang and Liu, Liheng and Li, Xiao and Chen, Min and Hong, Zhiliang and Xu, Jiawei},
  journal={IEEE Journal of Solid-State Circuits (JSSC)}, 
  title={{A 97.3-dB SNR Bioimpedance AFE With -84-dB THD Segmented- $\Delta\Sigma$ M Sinusoidal Current Generator and Passing-Through Instrumentation Amplifier}}, 
  year={2025},
  volume={60},
  number={4},
  pages={1411-1422},
  doi={10.1109/JSSC.2024.3516040}}

@INPROCEEDINGS{Lee2020eit-cicc,
  author={Lee, Jaehyuk and Gweon, Surin and Lee, Kwonjoon and Um, Soyeon and Lee, Kyoung-Rog and Kim, Kwantae and Lee, Jihee and Yoo, Hoi-Jun},
  booktitle={IEEE Custom Integrated Circuits Conference (CICC)}, 
  title={{A 9.6 mW/Ch 10 MHz Wide-bandwidth Electrical Impedance Tomography IC with Accurate Phase Compensation for Breast Cancer Detection}}, 
  year={2020},
  volume={},
  number={},
  doi={10.1109/CICC48029.2020.9075950}}

@ARTICLE{Takhti2019bioz-tbiocas,
  author={Takhti, Mohammad and Odame, Kofi},
  journal={IEEE Transactions on Biomedical Circuits and Systems (TBioCAS)}, 
  title={{A Power Adaptive, 1.22-pW/Hz, 10-MHz Read-Out Front-End for Bio-Impedance Measurement}}, 
  year={2019},
  volume={13},
  number={4},
  pages={725-734},
  doi={10.1109/TBCAS.2019.2918262}}

@ARTICLE{Rezaee-Dehsorkh2022BioZ,
  author={Rezaee-Dehsorkh, Hamidreza and Ravanshad, Nassim and Shamsaki, Ali and Fakour, Maryam Rouhbakhsh and Aliparast, Peiman},
  journal={IEEE Transactions on Biomedical Circuits and Systems (TBioCAS)}, 
  title={{A Low-Power Single-Path Bio-Impedance Measurement System Using an Analog-to-Digital Converter for I/Q Demodulation}}, 
  year={2022},
  volume={},
  number={},
  doi={10.1109/TBCAS.2022.3213869}}

@ARTICLE{Yang2023neural-jssc,
  author={Yang, Xiaolin and Ballini, Marco and Sawigun, Chutham and Hsu, Wen-Yang and Weijers, Jan-Willem and Putzeys, Jan and Lopez, Carolina Mora},
  journal={IEEE Journal of Solid-State Circuits}, 
  title={{An AC-Coupled 1st-Order $\Delta-\Delta\Sigma$ Readout IC for Area-Efficient Neural Signal Acquisition}}, 
  year={2023},
  volume={58},
  number={4},
  pages={949-960},
  doi={10.1109/JSSC.2023.3234612}}

@book{sansen2006essential,
  author = {Willy Sansen},
  title = {{Analog Design Essentials}},
  year = {2006},
  note = {{Chapter 4: Noise performance of elementary transistor stages}},
  publisher={New York: Springer}}

@book{razavi2001design,
  author = {Behzad Razavi},
  title = {{Design of Analog CMOS Integrated Circuits}},
  year = {2017},
  edition={second},
  note = {{Chapter 7.4.2: Common-Gate Stage}},
  publisher={McGraw-Hill}}

@ARTICLE{Kim2023filt-casm,
  author={Kim, Kwantae and Liu, Shih-Chii},
  journal={IEEE Circuits and Systems Magazine}, 
  title={{Continuous-Time Analog Filters for Audio Edge Intelligence: Review on Circuit Designs}}, 
  year={2023},
  volume={23},
  number={2},
  pages={29-48},
  doi={10.1109/MCAS.2023.3267893}}

@INPROCEEDINGS{Zhong2022adc-a-sscc,
  author={Zhong, Yi and Jie, Lu and Sun, Nan},
  booktitle={IEEE Asian Solid-State Circuits Conference (A-SSCC)}, 
  title={{A 78.6 dB-SNDR 520mV\textsubscript{pp}-full-scale 620M$\Omega$-Z\textsubscript{in} 105dB-CMRR VCO-based Sensor Readout Circuit Using FVF-Based Gm-Input Structure}}, 
  year={2022},
  volume={},
  number={},
  pages={1-3},
  doi={10.1109/A-SSCC56115.2022.9980716}}

@ARTICLE{Assaderaghi1997TED,
  author={Assaderaghi, F. and Sinitsky, D. and Parke, S.A. and Bokor, J. and Ko, P.K. and Chenming Hu},
  journal={IEEE Transactions on Electron Devices}, 
  title={{Dynamic threshold-voltage MOSFET (DTMOS) for ultra-low voltage VLSI}}, 
  year={1997},
  volume={44},
  number={3},
  pages={414-422},
  doi={10.1109/16.556151}}

@ARTICLE{Pan2022BioZ,
  author={Pan, Qinjing and Qu, Tianxiang and Tang, Biao and Shan, Fei and Hong, Zhiliang and Xu, Jiawei},
  journal={IEEE Journal of Solid-State Circuits (JSSC)}, 
  title={{A 0.5-m$\Omega$/$\surd$Hz Dry-Electrode Bioimpedance Interface With Current Mismatch Cancellation and Input Impedance of 100 M$\Omega$ at 50 kHz}}, 
  year={2023},
  volume={58},
  number={6},
  pages={1735-1745},
  doi={10.1109/JSSC.2022.3199492}}

@ARTICLE{Zhang2023BioZ-JSSC,
  author={Zhang, Tan-Tan and Son, Hyunwoo and Zhao, Jianming and Heng, Chun-Huat and Gao, Yuan},
  journal={IEEE Journal of Solid-State Circuits (JSSC)}, 
  title={{A 26.6--119.3-$\mu$W 101.9-dB SNR Direct Digitization Bio-Impedance Readout IC}}, 
  year={2023},
  volume={58},
  number={9},
  pages={2619-2631},
  doi={10.1109/JSSC.2023.3269671}}

@ARTICLE{Kim2022KWS-JSSC,
  author={Kim, Kwantae and Gao, Chang and Graça, Rui and Kiselev, Ilya and Yoo, Hoi-Jun and Delbruck, Tobi and Liu, Shih-Chii},
  journal={IEEE Journal of Solid-State Circuits (JSSC)}, 
  title={{A 23-$\mu$W Keyword Spotting IC With Ring-Oscillator-Based Time-Domain Feature Extraction}}, 
  year={2022},
  volume={57},
  number={11},
  pages={3298-3311},
  doi={10.1109/JSSC.2022.3195610}}

@INPROCEEDINGS{Nauta2024ISSCC,
  author={Nauta, Bram},
  booktitle={IEEE International Solid-State Circuits Conference (ISSCC)}, 
  title={{Racing Down the Slopes of Moore’s Law}}, 
  year={2024},
  volume={67},
  number={},
  pages={16-23},
  doi={10.1109/ISSCC49657.2024.10454417}}

@INPROCEEDINGS{Kinget2015ISSCC,
  author={Kinget, Peter},
  booktitle={IEEE International Solid-State Circuits Conference (ISSCC) Digest of Technical Papers}, 
  title={{Short Course: Designing Ultra Low Voltage Analog and Mixed Signal Circuits}}, 
  year={2015},
  volume={TUTORIAL},
  number={},
  pages={1-29},
  doi={10.1109/ISSCC15167.2015.11005903}}

@ARTICLE{Song2015JSSC-Glucose,
  author={Song, Kiseok and Ha, Unsoo and Park, Seongwook and Bae, Joonsung and Yoo, Hoi-Jun},
  journal={IEEE Journal of Solid-State Circuits (JSSC)}, 
  title={{An Impedance and Multi-Wavelength Near-Infrared Spectroscopy IC for Non-Invasive Blood Glucose Estimation}}, 
  year={2015},
  volume={50},
  number={4},
  pages={1025-1037},
  doi={10.1109/JSSC.2014.2384037}}

@ARTICLE{JSSC22Choi,
  author={Choi, Woojun and Angevare, Jan and Park, Injun and Makinwa, Kofi A. A. and Chae, Youngcheol},
  journal={IEEE Journal of Solid-State Circuits (JSSC)}, 
  title={{A 0.9-V 28-MHz Highly Digital CMOS Dual-RC Frequency Reference With ±200 ppm Inaccuracy From -40 °C to 85 °C}}, 
  year={2022},
  volume={57},
  number={8},
  pages={2418-2428},
  doi={10.1109/JSSC.2021.3135939}}

@ARTICLE{Chatterjee2005JSSC,
  author={Chatterjee, S. and Tsividis, Y. and Kinget, P.},
  journal={IEEE Journal of Solid-State Circuits (JSSC)}, 
  title={{0.5-V analog circuit techniques and their application in OTA and filter design}}, 
  year={2005},
  volume={40},
  number={12},
  pages={2373-2387},
  doi={10.1109/JSSC.2005.856280}}

@ARTICLE{Steyaert1987JSSC-NEF,
  author={Steyaert, M.S.J. and Sansen, W.M.C.},
  journal={IEEE Journal of Solid-State Circuits (JSSC)}, 
  title={{A micropower low-noise monolithic instrumentation amplifier for medical purposes}}, 
  year={1987},
  volume={22},
  number={6},
  pages={1163-1168},
  doi={10.1109/JSSC.1987.1052869}}

@ARTICLE{Harrison2003JSSC,
  author={Harrison, R.R. and Charles, C.},
  journal={IEEE Journal of Solid-State Circuits (JSSC)}, 
  title={{A low-power low-noise CMOS amplifier for neural recording applications}}, 
  year={2003},
  volume={38},
  number={6},
  pages={958-965},
  doi={10.1109/JSSC.2003.811979}}

@ARTICLE{Yan2013TBioCAS,
  author={Yan, Long and Pettine, Julia and Mitra, Srinjoy and Kim, Sunyoung and Jee, Dong-Woo and Kim, Hyejung and Osawa, Masato and Harada, Yasunari and Tamiya, Kosei and Van Hoof, Chris and Yazicioglu, Refet Firat},
  journal={IEEE Transactions on Biomedical Circuits and Systems (TBioCAS)}, 
  title={{A 13 $\mu {\rm A}$ Analog Signal Processing IC for Accurate Recognition of Multiple Intra-Cardiac Signals}}, 
  year={2013},
  volume={7},
  number={6},
  pages={785-795},
  doi={10.1109/TBCAS.2013.2297353}}

@ARTICLE{Ha2019JSSC,
  author={Ha, Hyunsoo and Sijbers, Wim and Van Wegberg, Roland and Xu, Jiawei and Konijnenburg, Mario and Vis, Peter and Breeschoten, Arjan and Song, Shuang and Van Hoof, Chris and Helleputte, Nick Van},
  journal={IEEE Journal of Solid-State Circuits (JSSC)}, 
  title={{A Bio-Impedance Readout IC With Digital-Assisted Baseline Cancellation for Two-Electrode Measurement}}, 
  year={2019},
  volume={54},
  number={11},
  pages={2969-2979},
  doi={10.1109/JSSC.2019.2939077}}

\section{Biography Section}

\begin{IEEEbiography}[{\includegraphics[width=1in,height=1.25in,clip,keepaspectratio]{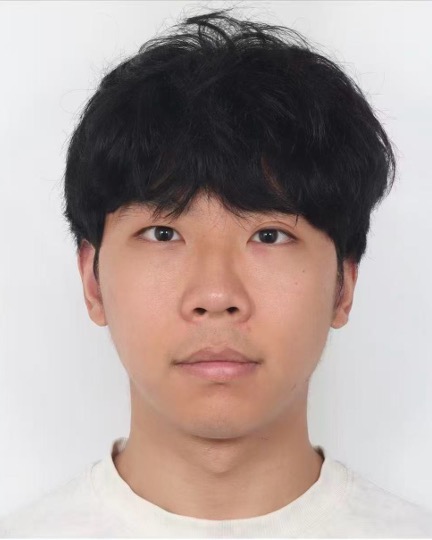}}]{Yu Xue}
(Student Member, IEEE) received the B.S. degree in electronic information engineering from Nanjing Tech University, Nanjing, China, in 2022. 
He is currently pursuing the M.S. degree in electronics and nanotechnology at Aalto University, Espoo, Finland, where he is working with the Tiny Systems and Circuits (TSirc) Research Group. 
His current research interests include low-noise and low-power instrumentation amplifiers for biomedical sensing applications.
\end{IEEEbiography}

\begin{IEEEbiography}[{\includegraphics[width=1in,height=1.25in,clip,keepaspectratio]{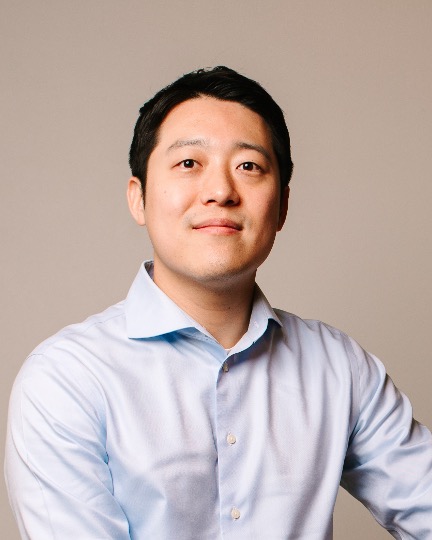}}]{Kwantae Kim}
(Senior Member, IEEE) received B.S., M.S., and Ph.D. degrees in the School of Electrical Engineering, KAIST, South Korea, in 2015, 2017, and 2021, respectively. He is an Assistant Professor at the Department of Electronics and Nanoengineering, School of Electrical Engineering, Aalto University, Finland.
He was a Visiting Student in 2020, and a Postdoctoral Researcher from 2021 to 2023, at the Institute of Neuroinformatics, University of Zurich and ETH Zurich, and an Established Researcher from 2023 to 2024, at the Department of Information Technology and Electrical Engineering, ETH Zurich, Switzerland. His research interests include analog/mixed-signal ICs and full-custom memory ICs for neuromorphic signal processing, biomedical sensors, and flexible electronics.
\end{IEEEbiography}

\end{document}